\documentclass[pra,twocolumn,floatfix]{revtex4}
                        \usepackage{graphicx}
                        \usepackage{dcolumn}
                        
\begin{document}

                        \def\be{\begin{equation}}
                        \def\ee{\end{equation}}
                        \def\ba{\begin{eqnarray}}
                        \def\ea{\end{eqnarray}}
                        \def\bas{\begin{eqnarray*}}
                        \def\eas{\end{eqnarray*}}


\title{A variational model for the delayed collapse of Bose Einstein condensates}

\author{Stavros Theodorakis and Stavros Athanasiou}
                        \affiliation{Physics Department, University of Cyprus,
P.O. Box 20537, Nicosia 1678, Cyprus}
                        \email{stavrost@ucy.ac.cy}
\date{\today}

\begin{abstract}
We present an action that can be used to study variationally the collapse of Bose Einstein condensates. This action is real, even though it includes dissipative terms. It adopts long range interactions between the atoms, so that there is always a stable minimum of the energy, even if the remaining number of atoms is above the number that in the case of local interactions is the critical one. The proposed action incorporates the time needed for the abrupt and delayed onset of collapse, yielding in fact its dependence on the scattering length. We show that the evolution of the condensate is equivalent to the motion of a particle in an effective potential. The particle begins its motion far from the point of stable equilibrium and it then proceeds to oscillate about that point. We prove that the resulting large oscillations in the shape of the wavefunction after the collapse have frequencies equal to twice the frequencies of the traps. Our results agree with the experimental observations. 
\end{abstract}

\maketitle

\vskip 0.3cm
\vskip 0.3cm

{\bf I. Introduction}

Two most intriguing aspects of the experimental results on the collapse of Bose Einstein condensates are its abrupt and delayed onset and then the survival of a remarkably stable remnant condensate with a constant number of atoms for more than one second\cite{Donley}. In these experiments the repulsive nature of the interactions between the atoms that were trapped in a magnetic trap was changed abruptly to attractive, resulting in the collapse of the condensate. A proposed explanation for the delayed onset of this collapse maintained that the condensate conserves initially the number of its atoms while shrinking in size. Thus the density gradually increases at the center of the condensate. When it becomes large enough, three-body recombination losses set in at the center, resulting thus in the expulsion of atoms from the condensate\cite{Ueda}.

The theoretical descriptions of the collapse involved the usual local Gross-Pitaevskii equation, augmented by a quintic dissipative term due to three body recombination processes\cite{Savage}. If one also includes the linear term describing the atomic feeding of the condensate from the surrounding nonequilibrium thermal cloud\cite{GPusers}, then one will end up with the generalized Gross-Pitaevskii equation

\ba
\label{GPeqn}
&&i\hbar\frac{\partial\psi}{\partial t}=-\frac{\hbar^{2}}{2m}\nabla^{2}\psi+U({\bf r})\psi-g|\psi|^{2}\psi+i\hbar\nu\psi\nonumber\\
&&-i\hbar\xi|\psi|^{4}\psi,
\ea
where $\nu$ and $\xi$ are real constant parameters. Here $U({\bf r})$ is the real harmonic potential of the trap, $\int|\psi|^{2} d^{3}r=N$ and $g=4\pi\hbar^{2}a/m$, $N$ being the number of atoms in the condensate and $a$ being the absolute value of the negative scattering length. This equation leads to the depletion rate of the condensate:

\be
\label{attrition}
\frac{dN}{dt}=2\nu N-2\xi\int|\psi|^{6} d^{3}r.
\ee

Hence the number of condensed atoms will become constant only if the right hand side becomes zero for long times. This involves $\psi$ having a steady profile, corresponding to the compensation for the three-body recombination losses by a steady influx of thermal noncondensed atoms from the surrounding cloud. In particular, the condensate density at the center of the trap and the width of the wavefunction will have to be constant at long times. This disagrees with the experimental results\cite{Donley}, which show that the width of the remnant condensate keeps oscillating in time, while the number of condensed atoms remains constant. Thus the longevity of the remnant condensate for times of the order of one second cannot be explained this way.

A way out seems to be offered by the realization that if the attractive interactions between the atoms are long-ranged and nonsingular, then the condensate cannot collapse\cite{Parola}. In fact, there are two energy minima below the critical point: a large width metastable anisotropic condensate, which disappears at the critical point, and a small width isotropic stable remnant, which continues to exist for values of $a$ much higher than the critical one, such as the ones used in the experiments\cite{Donley}.

Indeed, the existence of the high density and small width minimum of the energy is absolutely vital for the observation of a long-lived remnant. It explains, in fact, why the remnant has been often observed for numbers of atoms far greater than the number corresponding to the critical point. It is not sufficient though, as can be seen from Eq.~(\ref{attrition}). The nonlocal nature of the attractive interactions does alter the cubic term of Eq.~(\ref{GPeqn}), but no cubic terms appear in Eq.~(\ref{attrition}). Therefore the paradox persists.

This paper will propose a mechanism for resolving this contradiction between the experimental results and the formalism of the generalized Gross-Pitaevskii equation. In doing so, it will also give a detailed variational description of the collapse.

We shall adopt a phenomenological time-dependent dissipative term that can reproduce the observed evolution of the number of atoms of the condensate, including the delayed onset of the collapse. In fact, we shall be able to describe in detail the collapse of the Bose-Einstein condensate, finding in addition the dependence of the time of collapse on the scattering length. We shall achieve this by noting that the terms on the right hand side of Eq.~(\ref{attrition}) should vanish for long times, irrespective of what the width of the wavefunction looks like. This can only be achieved if both coefficients of the dissipative terms of Eq.~(\ref{GPeqn}) vanish at long times. In fact, both dissipative terms in Eq.~(\ref{GPeqn}) can be modelled by a phenomenological linear term with an imaginary time-dependent coefficient that will effectively encompass both the influx of noncondensed atoms and the dissipative losses. These will end up balancing each other, resulting thus eventually in the vanishing of this coefficient.

A variational  description will enable us to examine fully the behavior of the condensate. In order to do this though, we shall need to formulate a real action that leads to the desired generalized Gross-Pitaevskii equation, including the dissipative terms. We shall write down precisely such an action, enabling us to find the evolution of the wavefunction profile. This action will be used in the next section for determining the evolution of the number of atoms and comparing it with the observations.

In Section III we use a simple trial wavefunction for the case of an isotropic trap in order to find the effective potential and the critical point, as well as to explain the persistent oscillations of the remnant condensate. We also study the perturbations around the critical point and we find the dependence of the time of collapse on the scattering length. In Section IV we repeat these calculations using an anisotropic trial wavefunction and we compare the results to the experimental observations. Section V summarizes our results.

{II. \bf A real action}

We shall adopt the generalized Gross-Pitaevskii equation

\ba
\label{GPeqngeneral}
&&i\hbar\frac{\partial\Psi}{\partial t}=-\frac{\hbar^{2}}{2m}\nabla^{2}\Psi+\frac{m\omega_{\rho}^{2}\rho^{2}}{2}\Psi+\frac{m\omega_{z}^{2}z^{2}}{2}\Psi+i\hbar\nu(t)\Psi\nonumber\\
&&-\frac{4\pi a\hbar^{2}}{m}\int V({\bf r}-{\bf r^{\prime}})|\Psi({\bf r^{\prime}},t)|^{2}\Psi({\bf r},t)d^{3}r^{\prime}.
\ea

The magnetic trap is cylindrically symmetric with frequencies $\omega_{\rho}$ and $\omega_{z}$. For the long range interaction we assume that $\int V({\bf r})d^{3}r=1$, so that in the limit of zero range it will reduce to a Dirac delta function, turning then the nonlocal term into the standard cubic local term of Eq.~(\ref{GPeqn}). The time dependent coefficient $\nu(t)$ is complex, hence the whole equation is a dissipative one.

We can make this equation dimensionless\cite{Gammal}, if we measure $\Psi$ in units of $\sqrt{N_{0}/d^{3}}$, distances in units of $d$, times in units of $1/\omega$, $V$ in units of $1/d^{3}$ and $\nu(t)$ in units of $\omega$, where $d=\sqrt{\hbar/(2m\omega)}$ and $\omega=(\omega_{z}\omega_{\rho}^{2})^{1/3}$. Thus Eq.~(\ref{GPeqngeneral}) takes the form

\ba
\label{GPdimless}
&&i\frac{\partial\Psi}{\partial t}=-\nabla^{2}\Psi+\frac{\lambda^{-2/3}\rho^{2}}{4}\Psi+\frac{\lambda^{4/3}z^{2}}{4}\Psi+i\nu(t)\Psi\nonumber\\
&&-8\pi k\sqrt{2}\int V({\bf r}-{\bf r^{\prime}})|\Psi({\bf r^{\prime}},t)|^{2}\Psi({\bf r},t)d^{3}r^{\prime},
\ea

where $\lambda=\omega_{z}/\omega_{\rho}$, $k=N_{0}a/\ell_{0}$, $\ell_{0}=\sqrt{\hbar/(m\omega)}$. Here $N_{0}$ is the initial value of the number of atoms $N(t)$ in the condensate, so that $\int|\Psi|^{2}d^{3}r=N(t)/N_{0}=n(t)$.

If we multiply Eq.~(\ref{GPdimless}) by $\Psi^{*}$, subtract from the resulting expression its corresponding complex conjugate and then integrate over all space, we shall obtain the relation

\be
\label{derofn}
\frac{dn}{dt}=2n(t)Re\nu(t)
\ee
and hence
\be
\label{nexpression}
n(t)=e^{\int_{0}^{t}2Re\nu(\tau)d\tau}.
\ee

Thus dissipation will take place only as long as the real part of $\nu(t)$ is nonzero. In this formulation, the details of the nonlocal interactions do not affect directly the evolution of $n(t)$.

Eq.~(\ref{GPdimless}) minimizes the action $\int L_{0}(t)\,dt-i\int\nu(t)|\Psi|^{2}d^{3}r dt$, where

\ba
\label{L0}
&&L_{0}(t)=\int\,\,d^{3}r\Bigl(\frac{i}{2}\Psi^{*}\frac{\partial\Psi}{\partial t}-\frac{i}{2}\Psi\frac{\partial\Psi^{*}}{\partial t}-|\nabla\Psi|^{2}\nonumber\\
&&-\lambda^{-2/3}\frac{\rho^{2}}{4}|\Psi|^{2}-\lambda^{4/3}\frac{z^{2}}{4}|\Psi|^{2}\Bigr)\nonumber\\
           &&+4\pi k\sqrt{2}\int\,d^{3}r\,d^{3}r^{\prime}\,|\Psi({\bf r,t})|^{2} V({\bf r}-{\bf r^{\prime}})|\Psi({\bf r^{\prime},t})|^{2}
            \ea
						
The piece $\int L_{0}dt$ of the action is real, but the term $-i\int\nu(t)|\Psi|^{2}d^{3}r dt$ is not. Thus we cannot use this action in a variational calculation, since it cannot be minimized. We shall use a modified action instead, similar to the one used in the problem of a damped harmonic oscillator\cite{Bateman}:

\ba
\label{S}
&&S=\int\,e^{-\int_{0}^{t}2Re\nu(\tau)d\tau}L_{0}(t)dt\nonumber\\
&&+\int\,\,d^{3}r\,dt\,e^{-\int_{0}^{t}2Re\nu(\tau)d\tau}Im\nu(t)|\Psi|^{2}.
\ea

We can easily verify that the functional differentiation of $S$ with respect to $\Psi^{*}$ yields Eq.~(\ref{GPdimless}). It is the real action of Eq.~(\ref{S}) the action on which we shall base all our calculations.

\begin{figure}[t]
\vskip 0.3cm
                        \includegraphics[width=0.49\textwidth]{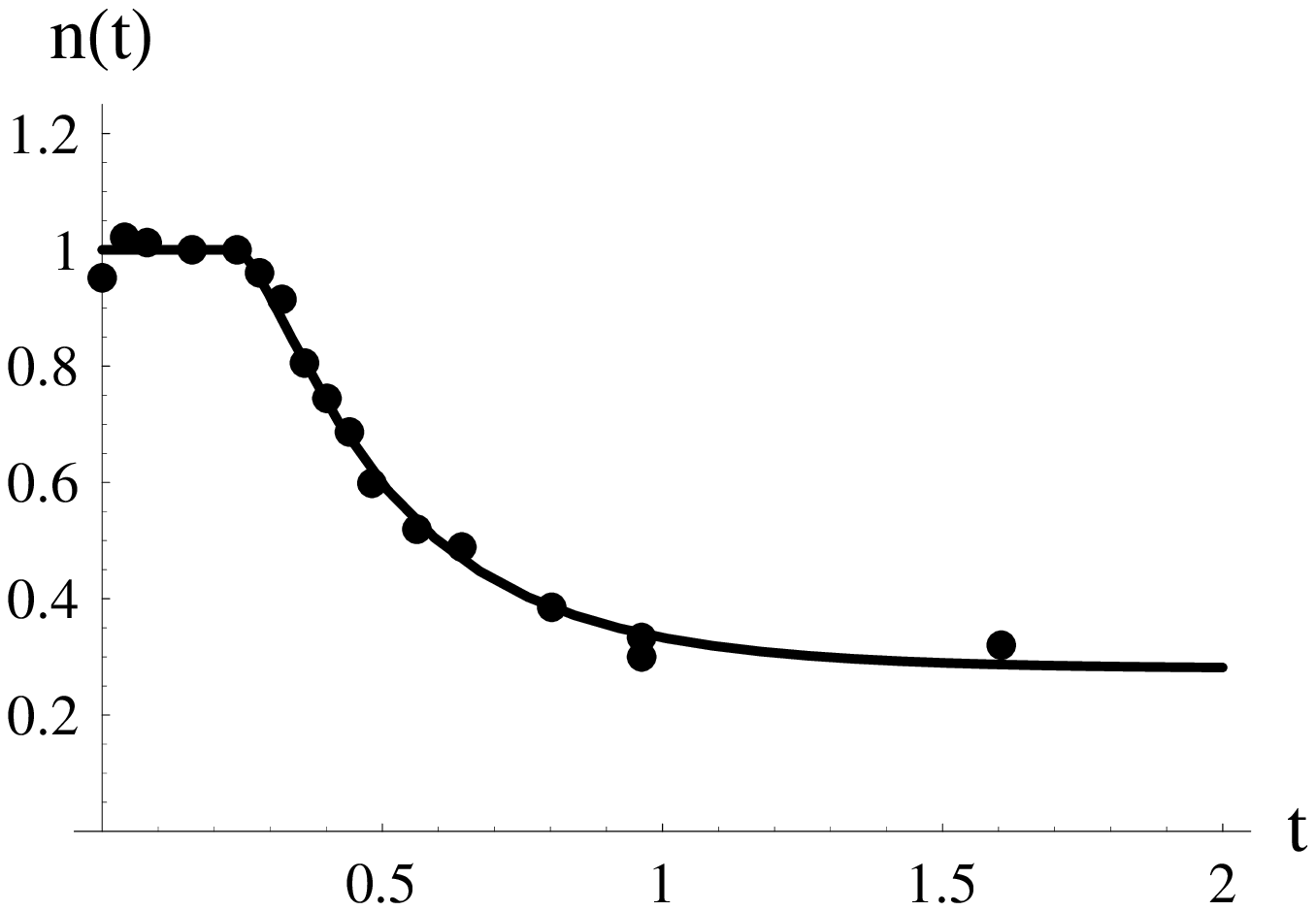}
                        \caption{\label{fig1}The number of atoms $n(t)=N(t)/N_{0}$ in the condensate  versus time for $a=36a_{0}$ and $N_{0}=16000$. The continuous line is the expression of Eq.~(\ref{gexpression}) when $t_{c}=0.2407$, $t_{0}=
0.2745$ and $\nu_{0}=2.6173$, while the points are the experimental data of Figure 4.2 of Ref.\cite{Claussen}.}
                        \end{figure}

In the experiment\cite{Donley} the scattering length is almost zero at time $t=0$, the initial condensate wavefunction being the harmonic oscillator ground state. It then jumps to 36 Bohr radii\cite{Claussen} within 0.1 msec. The attractive interaction is thus switched on suddenly and the condensate will absorb almost instantaneously any noncondensed atoms happen to be around it. In the experiment the condensate contained 97 percent of the total number of atoms. The initial sweeping of the surrounding noncondensed atoms is the reason for the slight initial positive slope of $n(t)$, seen in Figure~\ref{fig1}. Nonetheless, the maximum number of 16000 atoms is reached almost instantaneously. This number remains constant while the condensate is shrinking, till the central density becomes large enough to enhance dramatically the expulsion of atoms due to the three-body recombination losses. Thus $n(t)$ starts decreasing, till it reaches eventually an asymptotic value. The reason the collapse stops is the long range attraction of each atom by its outlying neighbors, which provides a vital enhancement to the quantum pressure and balances the attraction of the trap and of the central atoms. This balance is manifested by the ubiquitous existence of a stable isotropic small width minimum of the energy, irrespective of the value of $n(t)$, and can result in the appearance of remnants with a number of atoms far greater than the critical number.

We can describe all this behavior by choosing for the phenomenological parameter $\nu(t)$ the simplest form that is consistent with the experimental observations:

\be
\label{nuoft}
\nu(t)=-\nu_{0}\sqrt{\frac{t-t_{c}}{t_{0}}}e^{-(t-t_{c})/t_{0}}
\ee

Here $t_{c}$ is the collapse time, i.e. the time during which there is no change in the total number of condensate atoms. Indeed, if $0<t<t_{c}$, the real part of the $\nu(t)$ given by Eq.~(\ref{nuoft}) is zero. Correspondingly, Eq.~(\ref{nexpression}) indicates that the number of atoms remains constant during this time and equal to 1. Furthermore, the vanishing of $\nu(t)$ at long times implies according to Eq.~(\ref{derofn}) that the number of atoms is constant in that limit. This is what we mean by remnant condensate.

If $t>t_{c}$, then Eq.~(\ref{nexpression}) gives

\ba
\label{gexpression}
&&n(t)=Exp\Bigl[2e^{-(t-t_{c})/t_{0}}\nu_{0}\sqrt{t_{0}}\sqrt{t-t_{c}}\nonumber\\
&&-\sqrt{\pi}\nu_{0}t_{0}erf(\sqrt{(t-t_{c})/t_{0}})\Bigr]
\ea

We have fitted this expression for $n(t)$ to the data of Figure 4.2 of Ref.\cite{Claussen}, as shown in Figure~\ref{fig1}. The fit gives the dimensionless values $t_{c}=0.2407$, $t_{0}=
0.2745$ and $\nu_{0}=2.6173$. The initial number of atoms is $N_{0}=16000$, while $a$ rises within 0.1 msec from 0 to a constant value of $36a_{0}$, giving thus $k=9.98$. The final (asymptotic) value of $n(t)$ is 0.280, corresponding to an asymptotic value of 2.79 for $n(t)k$, still much greater than the corresponding critical value of 0.55\cite{Gammal}. The trap frequencies are $\omega_{z}=2\pi\times$6.8 Hz and $\omega_{\rho}=2\pi\times$17.5 Hz, with $d=2.16\,\mu m$.

\begin{figure}[t]
\vskip 0.3cm
                        \includegraphics[width=0.49\textwidth]{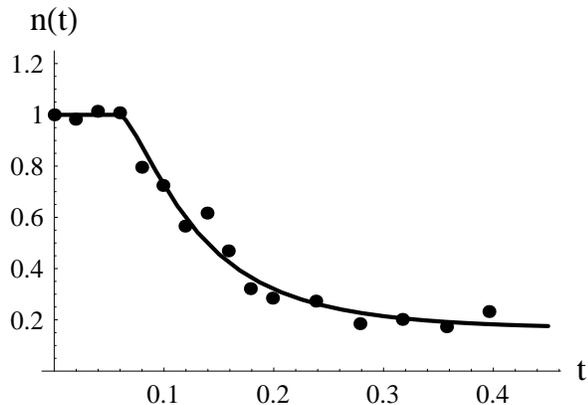}
                        \caption{\label{fig2}The number of atoms $n(t)=N(t)/N_{0}$ in the condensate  versus time for $a=20a_{0}$ and $N_{0}=40000$. The continuous line is the expression of Eq.~(\ref{gexpression}) when $t_{c}=0.0595$, $t_{0}=
0.0885$ and $\nu_{0}=11.4536$, while the points are the experimental data of Figure 1b of Ref.\cite{Altin}.}
                        \end{figure}

												We have also fitted the expression of Eq.~(\ref{gexpression}) to the data of Figure 1b of Ref.\cite{Altin}, as shown in Figure~\ref{fig2}. The fit gives the dimensionless values $t_{c}=0.0595$, $t_{0}=
0.0885$ and $\nu_{0}=11.4536$. The initial number of atoms in this case is $N_{0}=40000$, while $a$ rises almost instantaneously from 0 to a constant value of $20a_{0}$, giving thus $k=21.80$. The final (asymptotic) value of $n(t)$ is 0.166, corresponding to an asymptotic value of 3.61 for $n(t)k$, still much greater than the critical value of 0.557 for $k$\cite{Gammal}. The mean trap frequency is $\omega=2\pi\times$31.58 Hz, $\lambda=2.175$ and $d=1.37\,\mu m$.

\begin{figure}[t]
\vskip 0.3cm
                        \includegraphics[width=0.49\textwidth]{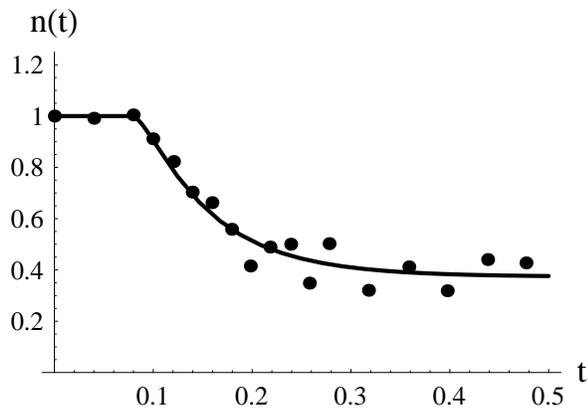}
                        \caption{\label{fig3}The number of atoms $n(t)=N(t)/N_{0}$ in the condensate  versus time for $a=8.4a_{0}$ and $N_{0}=40000$. The continuous line is the expression of Eq.~(\ref{gexpression}) when $t_{c}=0.0797$, $t_{0}=
0.0688$ and $\nu_{0}=8.070$, while the points are the experimental data of Figure 3 of Ref.\cite{Altin}.}
                        \end{figure}

												Finally, we have fitted the expression of Eq.~(\ref{gexpression}) to the data of Figure 3 of Ref.\cite{Altin}, as shown in Figure~\ref{fig3}. The fit gives the dimensionless values $t_{c}=0.0797$, $t_{0}=
0.0688$ and $\nu_{0}=8.070$. The initial number of atoms in this case is $N_{0}=40000$, while $a$ rises almost instantaneously from 0 to a constant value of $8.4a_{0}$, giving thus $k=9.15$. The final (asymptotic) value of $n(t)$ is 0.3737, corresponding to an asymptotic value of 3.42 for $n(t)k$, still much greater than the critical value of 0.557 for $k$\cite{Gammal}. The mean trap frequency is $\omega=2\pi\times$31.58 Hz and $d=1.37\,\mu m$.

												In all three of these cases the final value of $n(t)k$ is much greater than the critical value of 0.55. This particular value corresponds to the collapse in the local case, where the condensate collapses to a singularity when $k$ becomes large enough. The long range interactions generate however a stable isotropic minimum of the effective potential, the remnant condensate, so the concept of collapsing to a point singularity becomes irrelevant, along with the value $k_{crit}$=0.55.

\vskip 0.3cm
\vskip 0.3cm

{III. \bf The isotropic trap}

We shall use Eq.~(\ref{S}) in order to study variationally the case of an isotropic trap with $\lambda=1$ and $\omega_{z}=\omega_{\rho}$. We shall be interested in values of $k$ much greater than $k_{crit}$, thus we expect the density to be acutely peaked at the center. We shall adopt thus an exponential trial wavefunction of $r=|{\bf r}|$, rather than a gaussian:

\be
\label{ekthetiko}
\Psi({\bf r},t)=\frac{\sqrt{n(t)}}{\sqrt{\pi}s(t)^{3/2}}e^{-r/s(t)+ic(t)r^{2}+iw(t)},
\ee
where $s(t)$, $c(t)$ and $w(t)$ are real. This wavefunction satisfies the relation $\int|\Psi|^{2}d^{3}r=n(t)$.

The initial wavefunction at $t=0$ is $(2\pi)^{-3/4}e^{-r^{2}/4}$, the ground state of the harmonic oscillator, in which the mean value of $r^{2}$ is 3. In contrast, the mean value of $r^{2}$ for our exponential trial wavefunction is $3s(0)^{2}$ at time $t=0$. We shall require our trial wavefunction to have initially the same width as the initial wavefunction. Hence $s(0)=1$. Furthermore, $c(0)=0$ since the initial wavefunction has no $ir^{2}$ term in the exponent.

We shall assume a particular form now for the long range interaction:

\be
\label{nonlocalV}
V({\bf r})=\frac{e^{-r/\ell}}{8\pi\ell^{3}}.
\ee

When we insert our trial wavefunction of Eq.~(\ref{ekthetiko}) into Eq.~(\ref{S}), we obtain the action

\ba
\label{Slater}
&&\int\,dt\Bigl(-3s(t)^{2}c^{\prime}(t)-\frac{3}{4}s(t)^{2}-\frac{1+12c(t)^{2}s(t)^{4}}{s(t)^{2}}\nonumber\\
&&+\frac{n(t)k(32\ell^{2}+10\ell s(t)+s(t)^{2})}{\sqrt{2}(2\ell+s(t))^{5}}\nonumber\\
&&-w^{\prime}(t)+Im\nu(t)\Bigr)
\ea

The last two terms do not contribute to the dynamics. The functional derivative of this action with respect to $c(t)$ gives $c(t)=s^{\prime}(t)/(4s(t))$. We insert this expression for $c(t)$ into Eq.~(\ref{Slater}), obtaining finally the effective action

\be
\label{Seff}
S_{eff}=\int\,dt\Bigl(\frac{3}{4}s^{\prime}(t)^{2}-U_{eff}\Bigr)
\ee
and the effective energy

\be
\label{Heff}
H_{eff}=\frac{3}{4}s^{\prime}(t)^{2}+U_{eff},
\ee

where
\ba
\label{Ueff}
&&U_{eff}=\frac{1}{s(t)^{2}}+\frac{3s(t)^{2}}{4}\nonumber\\
&&-\frac{k_{eff}(t)(32\ell^{2}+10\ell s(t)+s(t)^{2})}{\sqrt{2}(2\ell+s(t))^{5}},
\ea

and $k_{eff}(t)=n(t)k$. We see thus that the dynamics is determined by the instantaneous value $k_{eff}(t)$.

\begin{figure}[t]
\vskip 0.3cm
                        \includegraphics[width=0.49\textwidth]{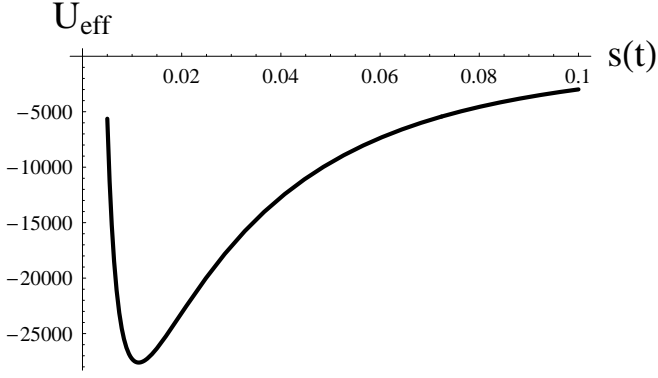}
                        \caption{\label{fig4}The effective potential $U_{eff}$ of Eq.~(\ref{Ueff}) when $k_{eff}(t)=9.98$ and $\ell=0.05$.}
                        \end{figure}

There is always at least one minimum of $U_{eff}$. For example, if $k_{eff}(t)=9.98$ and $\ell=0.05$ the wavefunction width is very small and the corresponding single minimum very deep (see Figure~\ref{fig4}). In fact, minimizing $U_{eff}$ for large $k_{eff}(t)$ and small $s(t)$ yields the width $s(t)=1.08943\ell^{4/3}/k_{eff}(t)^{1/3}$.

\begin{figure}[t]
\vskip 0.3cm
                        \includegraphics[width=0.49\textwidth]{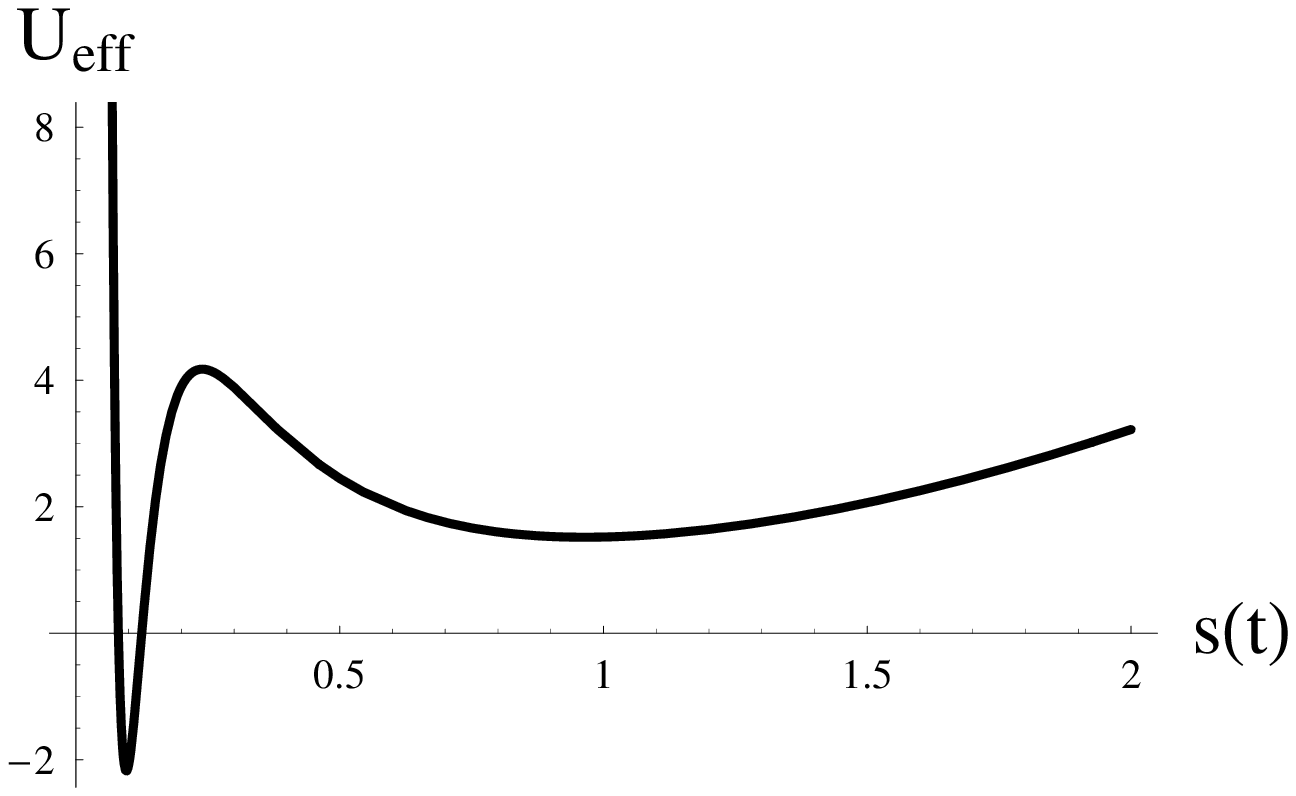}
                        \caption{\label{fig5}The effective potential $U_{eff}$ of Eq.~(\ref{Ueff}) when $k_{eff}(t)=0.33$ and $\ell=0.05$.}
                        \end{figure}

If however we take the example $k_{eff}(t)=0.33$ and $\ell=0.05$, there are two minima (see Figure~\ref{fig5}), a minimum with a large width and low density and the high density remnant with a small width. The remnant exists due to the long range interactions.

In the local case ($\ell=0$) the width of the remnant becomes zero, hence the remnant becomes a singularity. In that case the collapse is associated with the loss of stability of the unique minimum, the one with the large width. We can find the critical value $k_{crit}$ of $k_{eff}(t)$ in this local case by requiring that both the first and second derivatives of the effective potential $U_{eff}$ with respect to $s(t)$ vanish there. This happens when $s(t)=0.7186$ and $k_{crit}=0.542$. The exact value for the critical point of the local collapse in an isotropic trap is 0.5746\cite{Gammal}. Thus our variational model gives the correct critical value with an error of about 6 percent.

\begin{figure}[t]
\vskip 0.3cm
                        \includegraphics[width=0.49\textwidth]{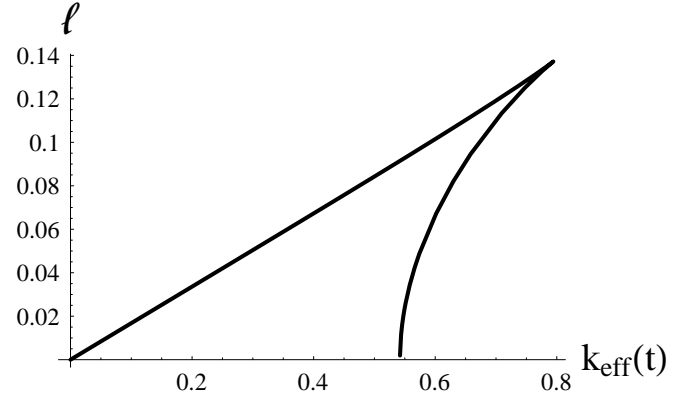}
                        \caption{\label{fig6} Within the triangular region shown above in the $k_{eff}(t)$-$\ell$ space, there are two possible minima of the effective potential. The borders of this region make both the first and the second derivative of $U_{eff}$ with respect to $s(t)$ equal to zero. Outside this region, there is only one minimum. The minimum on the right of the curved line corresponds to the remnant condensate.}
                        \end{figure}

Let us also examine the case $\ell=0.05$. The first and second derivatives of $U_{eff}$ with respect to $s(t)$ now become simultaneously zero for the values ($s(t)$, $k_{eff}(t)$) =(0.1486, 0.297) and (0.7013, 0.575). If we start with a zero value of $k_{eff}(t)$ and then increase it, we shall have initially a large width minimum, then at $k_{eff}(t)=0.297$ a second minimum with narrow width appears, and then at $k_{eff}(t)=0.575$ the large width minimum disappears, leaving only the remnant with the narrow width as a possible state. We can see in fact the regions of existence of one or two possible minima in the ($k_{eff}(t)$, $\ell$) graph of Figure~\ref{fig6}. Within the triangular region shown in the $k_{eff}(t)$-$\ell$ space of that figure, there are two possible minima of the effective potential. The borders of this region render both the first and the second derivative of $U_{eff}$ with respect to $s(t)$ equal to zero. Outside this region, there is only one minimum. The minimum on the right of the curved line corresponds to the remnant condensate.

We can explore further the dynamics of the condensate close to the point of collapse. Let us assume that the effective potential has a minimum at $s(t)=s_{0}$ for a given set of $\ell$ and $k_{eff}(t)$. We can expand the effective potential around $s_{0}$, obtaining

\be
\label{expansion}
U_{eff}(s(t))\approx\,U_{eff}(s_{0})+B(k_{eff}(t))(s(t)-s_{0})^2/2.
\ee
For a given value of $\ell$ there is a value $k_{crit}$ which is on the curved border of the triangular region of Figure~\ref{fig6}. At this value the first and second derivatives of $U_{eff}$ become zero and the large width minimum ceases to exist. Hence $B(k_{crit})=0$. For values of $k_{eff(t)}$ above $k_{crit}$ $B(k_{eff}(t))$ will be negative. We may expand it around $k_{crit}$ then and obtain

\be
\label{expansion2}
B(k_{eff}(t))\approx\,-(k_{eff}(t)-k_{crit})\Omega^{2}
\ee

Thus the action becomes

\ba
\label{Sapprox}
&&\int\,dt\Bigl(\frac{3}{4}s^{\prime}(t)^{2}-U_{eff}(s_{0})\nonumber\\
&&+(k_{eff}(t)-k_{crit})\Omega^{2}(s(t)-s_{0})^2/2\Bigr)
\ea

The corresponding equation of motion is

\be
\label{eqn}
\frac{3}{2}s^{\prime\prime}(t)=(k_{eff}(t)-k_{crit})\Omega^{2}(s(t)-s_{0})
\ee
For times before $t_{c}$ we expect $k_{eff}(t)=k$. The solutions of this equation involve then exponentials of the form $e^{t/\tau}$, where the quantity $\tau=\sqrt{\frac{3}{2(k-k_{crit})\Omega^{2}}}$ indicates the time needed for the manifestation of the instability. It is, in other words, essentially the time of collapse. We expect therefore the time of collapse $t_{c}$ to be proportional to $(k-k_{crit})^{-1/2}$, a conclusion that can also be reached by alternative arguments\cite{Calzetta}. This collapse time determines the beginning of the collapse towards the small width minimum. 

\begin{figure}[t]
\vskip 0.3cm
                        \includegraphics[width=0.49\textwidth]{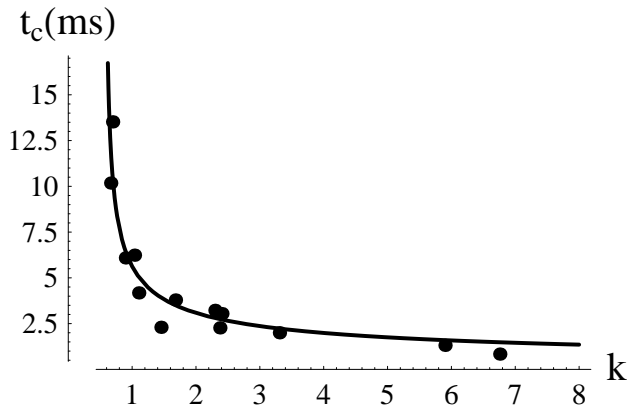}
                        \caption{\label{fig7}The collapse time (in msec) as a function of $k$, for $N_{0}=6000$ and $d=2.16\,\mu m$. The experimental points are taken from Figure 4.3 of Ref.\cite{Claussen} and fit quite well to the  function $f_{0}/\sqrt{k-k_{c}}$, where $f_{0}=3.6837$ and $k_{c}=0.5715$.}
                        \end{figure}

We can test this prediction by comparing it with the experimental data. The experiments used an anisotropic trap, but $t_{c}$ varies like $(k-k_{crit})^{-1/2}$ in that case too, as we shall see in the next section. In Figure~\ref{fig7} we show the collapse time $t_{c}$ (in msec) from Figure 4.3 of Ref.\cite{Claussen}, for which $N_{0}=6000$ and $d=2.16\,\mu m$. This data is fitted to a function of the form $f_{0}/\sqrt{k-k_{c}}$, where $f_{0}=3.6837$ and $k_{c}=0.5715$. We see that the agreement is quite good. We note that if we were to require the value $k=0.5715$ to be the point of collapse in this isotropic case, i.e. a minimum of $U_{eff}$ to disappear at this value of $k_{eff}(t)$, then we would need the minimum to occur at $s(t)=0.703$ and $\ell$ would need to be equal to 0.047.  

\begin{figure}[t]
\vskip 0.3cm
                        \includegraphics[width=0.49\textwidth]{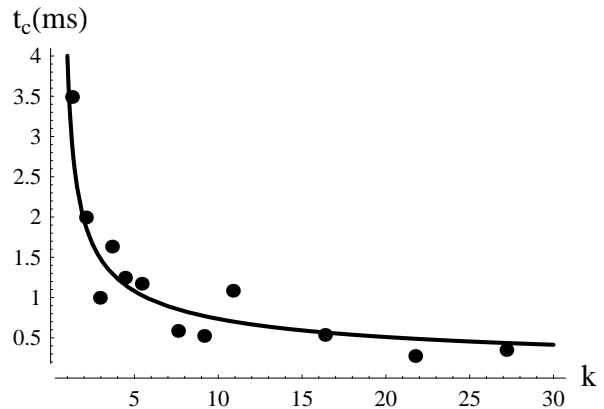}
                        \caption{\label{fig8}The collapse time (in msec) as a function of $k$, for $N_{0}=40000$ and $d=1.37\,\mu m$. The experimental points are taken from Figure 2 of Ref.\cite{Altin} and fit quite well to the  function $f_{0}/\sqrt{k-k_{c}}$, where $f_{0}=2.2411$ and $k_{c}=0.6865$.}
                        \end{figure}

In Figure~\ref{fig8} we show the collapse time $t_{c}$ (in msec) from Figure 2 of Ref.\cite{Altin}, for which $N_{0}=40000$ and $d=1.37\,\mu m$. This data is fitted to a function of the form $f_{0}/\sqrt{k-k_{c}}$, where $f_{0}=2.2411$ and $k_{c}=0.6865$. We see that the agreement is again quite good. We note that if we were to require the value $k=0.6865$ to be the point of collapse in this isotropic case, i.e. a minimum of $U_{eff}$ to disappear at this value of $k_{eff}(t)$, then we would need the minimum to occur at $s(t)=0.640$ and $\ell$ would need to be equal to 0.105.

\begin{figure}[t]
\vskip 0.3cm
                        \includegraphics[width=0.49\textwidth]{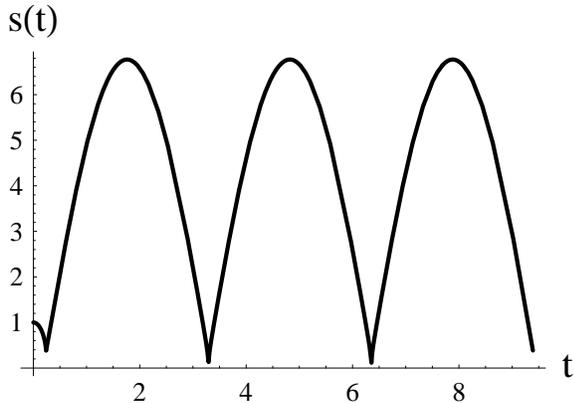}
                        \caption{\label{fig9}The oscillations of $s(t)$ as a function of time. Here $n(t)$ is equal to 1 if $0<t<t_{c}$, but it is given by Eq.~(\ref{gexpression}) for later times. We adopt the values $k=9.98$, $\ell=0.05$, $t_{c}=0.2407$, $t_{0}=0.2745$ and $\nu_{0}=2.6173$.}
                        \end{figure}

We can find the time evolution of the condensate by solving the Euler-Lagrange equation for the action of Eq.~(\ref{Seff}). We shall do so in fact for the data of Figure 4.2 of Ref.\cite{Claussen}, shown in Figure~\ref{fig1}. We shall adopt the values $k=9.98$, $\ell=0.05$, $t_{c}=0.2407$, $t_{0}=0.2745$ and $\nu_{0}=2.6173$. We shall assume that the initial value of $k_{eff}(t)$ is 9.98, since the value of $a$ is shifted almost instantaneously from 0 to 36$a_{0}$. The initial values $s(0)$ and $s^{\prime}(0)$ are then 1 and 0 respectively. In the interval (0,$t_{c}$) we have $n(t)=1$, but for later times it is given by Eq.~(\ref{gexpression}). The resulting numerical solution of the Euler-Lagrange equation for the action of Eq.~(\ref{Seff}) is shown in Figure~\ref{fig9}.

We see that the corresponding oscillations are huge and persistent. The reason for this can be understood if we look at Figure~\ref{fig4}. The action of Eq.~(\ref{Seff}) is the action of a particle moving in the effective potential $U_{eff}$. The particle starts at rest at the point $s[0]=1$, at the right edge of the deep potential well of this figure. It then accelerates towards the minimum and passes it, overshooting till it reaches a stopping point at a value of $s(t)=0.0047$ well beyond the minimum. Finally, it moves in the opposite direction, completing thus a full oscillation.

We can find the approximate form of $s(t)$ in these oscillations. For large values of $s(t)$ the Euler-Lagrange equation for Eq.~(\ref{Seff}) becomes

\be
\label{Euler}
\frac{3(s(t)+s^{\prime\prime}(t))}{2}=\frac{2}{s(t)^{3}}
\ee

The solution of this differential equation is

\be
\label{diffsol}
s(t)=\sqrt{\sqrt{\frac{4}{3}+\delta^{2}}+\delta\cos(2t-2t_{1})},
\ee
where $\delta$ and $t_{1}$ are integration constants. For the oscillation of Figure~\ref{fig9}, in which a maximum occurs at $t=1.7616$ and $s(t)=6.77054$, these constants take the values $\delta=22.9$ and $t_{1}=1.7616$. 
Eq.~(\ref{diffsol}) shows clearly that the width oscillates at the frequency $2\omega$ in an isotropic trap, irrespective of the value of $k_{eff}(t)$.

\vskip 0.3cm
\vskip 0.3cm

{IV. \bf The anisotropic trap}

We shall now study variationally the case of an anisotropic trap. Here we shall use a gaussian wavefunction of the cylindrical coordinates $\rho$ and $z$, since an exponential wavefunction would involve very complicated integrals:

\ba
\label{gaussian}
&&\Psi(\rho,z,t)=\frac{\sqrt{n(t)}}{\pi^{3/4}s1(t)\sqrt{s2(t)}}\times\nonumber\\
&&e^{-\frac{\rho^{2}}{2s_{1}(t)^{2}}-\frac{z^2}{2s_{2}(t)^{2}}+ic_{1}(t)\rho^{2}+ic_{2}(t)z^{2}+iw(t)},
\ea
where $s_{1}(t)$,  $s_{2}(t)$, $c_{1}(t)$, $c_{2}(t)$ and $w(t)$ are real. This wavefunction satisfies the relation $\int|\Psi|^{2}d^{3}r=n(t)$.
 
The initial wavefunction is $(2\pi)^{-3/4}e^{-\rho^{2}/(4\lambda^{1/3})-\lambda^{2/3}z^{2}/4}$, the ground state of the anisotropic harmonic oscillator. Hence $s1(0)=\sqrt{2}\lambda^{1/6}$ and $s_{2}(0)=\sqrt{2}\lambda^{-1/3}$. Furthermore, $c_{1}(0)=c_{2}(0)=0$ since the initial wavefunction has no $i\rho^{2}$ and $iz^{2}$ terms in the exponent.

We shall assume a gaussian form now for the long range interaction:

\be
\label{nonlocalV}
V(\rho,z)=\frac{e^{-\rho^{2}/\ell^{2}-z^{2}/\ell^{2}}}{\pi^{3/2}\ell^{3}}.
\ee

When we insert our trial wavefunction of Eq.~(\ref{gaussian}) into Eq.~(\ref{S}), we obtain the action

\ba
\label{Slateraniso}
&&\int\,dt\Bigl(-\frac{1}{s_{1}(t)^{2}}-\frac{1}{2s_{2}(t)^{2}}\nonumber\\
&&-s_{1}(t)^{2}\bigl(\frac{\lambda^{-2/3}}{4}+4c_{1}(t)^{2}+c_{1}^{\prime}(t)\bigr)\nonumber\\
&&-\frac{s_{2}(t)^{2}}{8}\bigl(\lambda^{4/3}+16c_{2}(t)^{2}+4c_{2}^{\prime}(t)\bigr)\nonumber\\
&&+\frac{4n(t)k\sqrt{2/\pi}}{(\ell^{2}+2s_{1}(t)^{2})\sqrt{\ell^{2}+2s_{2}(t)^{2}}}\nonumber\\
&&-w^{\prime}(t)+Im\nu(t)\Bigr)
\ea

The last two terms do not contribute to the dynamics. The functional derivatives of this action with respect to $c_{1}(t)$ and $c_{2}(t)$ give $c_{1}(t)=s_{1}^{\prime}(t)/(4s_{1}(t))$ and $c_{2}(t)=s_{2}^{\prime}(t)/(4s_{2}(t))$. We insert these expressions for $c_{1}(t)$ and $c_{2}(t)$ into Eq.~(\ref{Slateraniso}), obtaining finally the effective action

\be
\label{Seffaniso}
S_{eff}=\int\,dt\Bigl(\frac{1}{4}s_{1}^{\prime}(t)^{2}+\frac{1}{8}s_{2}^{\prime}(t)^{2}-U_{eff}\Bigr)
\ee
and the effective energy

\be
\label{Heffaniso}
H_{eff}=\frac{1}{4}s_{1}^{\prime}(t)^{2}+\frac{1}{8}s_{2}^{\prime}(t)^{2}+U_{eff},
\ee

where
\ba
\label{Ueffaniso}
&&U_{eff}=\frac{1}{s_{1}(t)^{2}}+\frac{s_{1}(t)^{2}}{4\lambda^{2/3}}\nonumber\\
&&+\frac{1}{2s_{2}(t)^{2}}+\frac{\lambda^{4/3}s_{2}(t)^{2}}{8}\nonumber\\
&&-\frac{4k_{eff}(t)\sqrt{2/\pi}}{(\ell^{2}+2s_{1}(t)^{2})\sqrt{\ell^{2}+2s_{2}(t)^{2}}}
\ea

and $k_{eff}(t)=n(t)k$. We see thus that the dynamics is determined by the instantaneous value $k_{eff}(t)$.

There is always at least one minimum of $U_{eff}$. For large $k_{eff}(t)$ the wavefunction widths are very small and the corresponding single minimum very deep. In fact, minimizing $U_{eff}$ for large $k_{eff}(t)$ and small $s_{1}(t)$ and $s_{2}(t)$ yields the widths of the remnant that is stabilised by the long range interactions:

\be
\label{small1}
s_{1}(t)\approx\frac{\sqrt{2}\ell^{5/4}\lambda^{1/6}}{(\ell^{5}+32\lambda^{2/3}\sqrt{2/\pi}k_{eff}(t))^{1/4}}
\ee
and
\be
\label{small2}
s_{2}(t)\approx\frac{\sqrt{2}\ell^{5/4}}{(\ell^{5}\lambda^{4/3}+32\sqrt{2/\pi}k_{eff}(t))^{1/4}}.
\ee
For large $k_{eff}(t)$ these widths are essentially equal, resulting in a spherical remnant.

In the local case ($\ell=0$) the width of the remnant becomes zero, hence the remnant becomes a singularity. In that case the collapse is associated with the loss of stability of the unique minimum, the one with the large width. We can find the critical value $k_{crit}$ of $k_{eff}(t)$ in this local case, for the case of $\lambda=6.8/17.5$ for example,  by requiring that there be a saddle point in the effective potential $U_{eff}$. This happens when $s_{1}(t)=0.8771$, $s_{2}(t)=0.9956$ and $k_{crit}=0.637$. The exact value for the critical point of the local collapse in such an anisotropic trap is 0.550\cite{Gammal}. Thus our variational model gives the correct critical value with an error of about 16 percent.

Let us also examine the case $\ell=0.05$ with $\lambda=6.8/17.5$. There are saddle points at ($s_{1}(t)$, $s_{2}(t)$, $k_{eff}(t)$) =(0.0707, 0.0707, 0.1095) and (0.876, 0.993, 0.6395). If we start with a zero value of $k_{eff}(t)$ and then increase it, we shall have initially an anisotropic minimum, then at $k_{eff}(t)=0.1095$ an isotropic minimum with equal narrow widths appears, and then at $k_{eff}(t)=0.6395$ the anisotropic minimum disappears, leaving only the spherical remnant with the narrow width as a possible state. In general, we always find that after collapse the remnant condensate wavefunction is spherically symmetric around the center of the trap, as shown above.

We can explore further the dynamics of the condensate close to the point of collapse, where the anisotropic minimum becomes a saddle point and thus the only viable minimum is the spherical remnant. Let us assume that the effective potential has a minimum at $s_{1}(t)=s_{10}$ and $s_{2}(t)=s_{20}$ for a given set of $\ell$, $\lambda$ and $k_{eff}(t)$. We can expand the effective potential around this minimum, obtaining

\ba
\label{expansionaniso}
&&U_{eff}\approx\,U_{0}+A_{11}(k_{eff}(t))(s_{1}(t)-s_{10})^2/2\nonumber\\
&&+A_{22}(k_{eff}(t))(s_{2}(t)-s_{20})^2/2\nonumber\\
&&+A_{12}(k_{eff}(t))(s_{1}(t)-s_{10})(s_{2}(t)-s_{20}).
\ea
For a given value of $\ell$ there is a value $k_{crit}$ at which the anisotropic minimum becomes a saddle point. At this value $A_{11}$ and $A_{22}$ are positive, while $A_{12}^{2}=A_{11}A_{22}$. For values of $k_{eff(t)}$ just above $k_{crit}$ the quantities $s_{1}(t)$ and $s_{2}(t)$ will increase exponentially as $e^{Wt}$. If we solve the resulting linear equations of motion, we shall find $W^{2}$ as a function of $k_{eff}(t)$:

\ba
\label{W}
&&W^{2}=-A_{11}-2A_{22}\nonumber\\
&&+\sqrt{A_{11}^{2}+8A_{12}^{2}-4A_{11}A_{22}+4A_{22}^2}.
\ea

At the critical point, where  $A_{12}^{2}=A_{11}A_{22}$, this $W$ becomes zero. Hence, if we expand $W$ just above $k_{crit}$, we shall obtain $W^{2}\approx W_{0}^{2}(k_{eff}(t)-k_{crit})$. Therefore the time $1/W$, the time that characterizes the manifestation of the instability, will be proportional to $1/\sqrt{k_{eff}(t)-k_{crit}}$. Consequently, the collapse time, which is of the same order of magnitude as $1/W$, will also vary like $1/\sqrt{k_{eff}(t)-k_{crit}}$, as was already shown to be the case for the isotropic trap and as the experimental observations in Figure~\ref{fig7} and Figure~\ref{fig8} indicate as well. 

\begin{figure}[t]
\vskip 0.3cm
                        \includegraphics[width=0.49\textwidth]{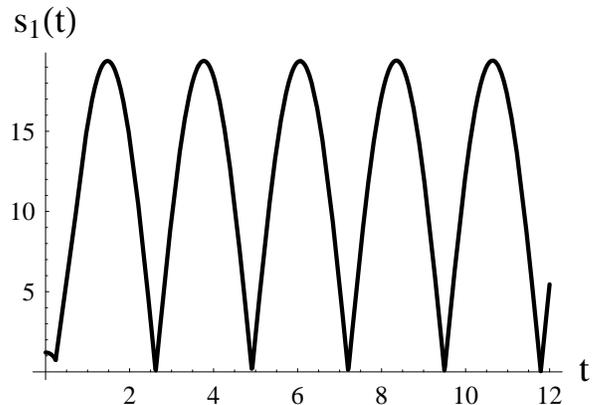}
                        \caption{\label{fig10}The oscillations of the radial width $s_{1}(t)$ as a function of time. Here $n(t)$ is equal to 1 if $0<t<t_{c}$, but it is given by Eq.~(\ref{gexpression}) for later times. We adopt the values $\lambda=6.8/17.5$, $k=9.98$, $\ell=0.05$, $t_{c}=0.2407$, $t_{0}=0.2745$ and $\nu_{0}=2.6173$.}
                        \end{figure}

\begin{figure}[t]
\vskip 0.3cm
                        \includegraphics[width=0.49\textwidth]{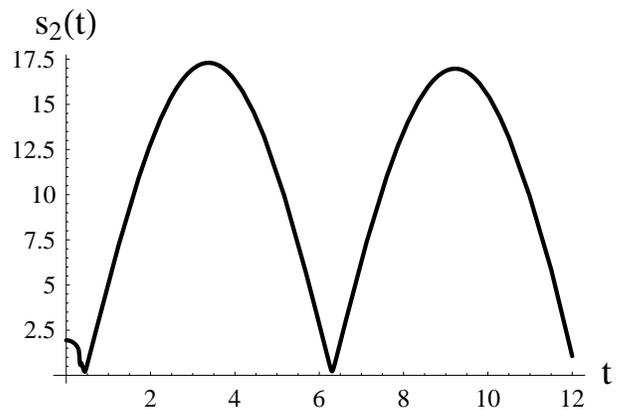}
                        \caption{\label{fig11}The oscillations of the axial width $s_{2}(t)$ as a function of time. Here $n(t)$ is equal to 1 if $0<t<t_{c}$, but it is given by Eq.~(\ref{gexpression}) for later times. We adopt the values $\lambda=6.8/17.5$, $k=9.98$, $\ell=0.05$, $t_{c}=0.2407$, $t_{0}=0.2745$ and $\nu_{0}=2.6173$.}
                        \end{figure}

We can find the time evolution of the condensate by solving the Euler-Lagrange equation for the action of Eq.~(\ref{Seffaniso}). We shall do so in fact for the data of Figure 4.2 of Ref.\cite{Claussen}, shown in Figure~\ref{fig1}. We shall adopt the values $\lambda=6.8/17.5$, $k=9.98$, $\ell=0.05$, $t_{c}=0.2407$, $t_{0}=0.2745$ and $\nu_{0}=2.6173$. We shall assume that the initial value of $k_{eff}(t)$ is 9.98, since the value of $a$ is shifted almost instantaneously from 0 to 36$a_{0}$. The initial values $s_{1}(0)$, $s_{2}(0)$, $s_{1}^{\prime}(0)$ and $s_{2}^{\prime}(0)$ are then 1.208, 1.938, 0 and 0 respectively. In the interval (0, $t_{c}$) we have $n(t)=1$, but for later times it is given by Eq.~(\ref{gexpression}). The resulting numerical solutions of the Euler-Lagrange equation for the action of Eq.~(\ref{Seffaniso}) are shown in Figure~\ref{fig10} and Figure~\ref{fig11}.

We see that the corresponding oscillations are again huge and persistent, just as in the isotropic case. The action of Eq.~(\ref{Seffaniso}) is the action of a particle moving in the effective potential $U_{eff}$. The particle starts at rest at the edge of the deep potential well. It then accelerates towards the spherical narrow width minimum and passes it, overshooting till it reaches a stopping point well beyond the minimum. Finally, it moves in the opposite direction, completing thus a full oscillation.

We can find the approximate form of the widths in these oscillations. For large values of $s_{1}(t)$ and $s_{2}(t)$ the Euler-Lagrange equations for Eq.~(\ref{Seffaniso}) become

\be
\label{Euler1}
\frac{s_{1}(t)}{\lambda^{2/3}}+s_{1}^{\prime\prime}(t)=\frac{4}{s_{1}(t)^{3}}
\ee
and
\be
\label{Euler2}
\lambda^{4/3}s_{2}(t)+s_{2}^{\prime\prime}(t)=\frac{4}{s_{2}(t)^{3}}
\ee

The solutions of these differential equations are

\be
\label{diffsol1}
s_{1}(t)=\sqrt{\sqrt{4\lambda^{2/3}+\delta_{1}^{2}}+\delta_{1}\cos(2\lambda^{-1/3}(t-t_{1}))},
\ee
and
\be
\label{diffsol2}
s_{2}(t)=\sqrt{\sqrt{4\lambda^{-4/3}+\delta_{2}^{2}}+\delta_{2}\cos(2\lambda^{2/3}(t-t_{2}))},
\ee

where $\delta_{1}$, $\delta_{2}$, $t_{1}$ and $t_{2}$ are integration constants. For the oscillations of Figure~\ref{fig10} and Figure~\ref{fig11}, in which a maximum for $s_{1}(t)$ occurs at $t=1.4754$ and $s_{1}(t)=19.3807$, the constants take the values $\delta_{1}=187.804$ and $t_{1}=1.4754$. Since a maximum for $s_{2}(t)$ occurs at $t=3.3754$ and $s_{2}(t)=17.300$, the constants take the values $\delta_{2}=149.619$ and $t_{2}=3.3754$. 
Eq.~(\ref{diffsol1}) and Eq.~(\ref{diffsol2}) show clearly that the widths $s_{1}(t)$ and $s_{2}(t)$ oscillate at the frequencies $2\lambda^{-1/3}\omega=2\omega_{\rho}$ and $2\lambda^{2/3}\omega=2\omega_{z}$ in an anisotropic trap, irrespective of the value of $k_{eff}(t)$. This is in fact what the experiments showed\cite{Donley}.

The single anisotropic gaussian trial function of Eq.~(\ref{gaussian}) is not very accurate. We can obtain a much more accurate trial wavefunction if we use the sum of two such anisotropic gaussians\cite{Andreas}. However, the resulting expressions are too lengthy and complicated. For the local case $\ell=0$ and for $\lambda=6.8/17.5$, the anisotropic minimum of the effective potential becomes a saddle point and the collapse occurs when $k_{eff}(t)=0.5533$, which agrees with the exact value\cite{Gammal}. For the case $\lambda=6.8/17.5$ and $k_{eff}(t)=0.5715$ of Figure~\ref{fig7},  a saddle point exists at $\ell=0.1277$. For the case $\lambda=2.175$ and $k_{eff}(t)=0.6865$ of Figure~\ref{fig8},  a saddle point exists at $\ell=0.3165$.

The range $\ell$ of the nonlocal interactions seems to be associated to the already existing lengths $d$ and $\ell_{B}$, where $\ell_{B}=\sqrt{\hbar/(eB)}$ is the magnetic length.  The scattering length is too small to be of the order of the interaction range for the nonlocal interactions.

The magnetic length is equal to $2.56\,\mu m/\sqrt{B}$ if $B$ is measured in Gauss. The scattering length $a$ near a Feshbach resonance is given by the expression

\be
\label{aFeshbach}
a(B)=a_{bg}\big(1-\frac{\Delta}{B-B_{p}}\big),
\ee

where $a_{bg}$ is the background scattering length, $\Delta$ is the resonance width and $B_{p}$ is the resonance centre, these quantities having the values\cite{error} $a_{bg}=-443\,a_{0}$, $\Delta=10.71\,G$ and $B_{p}=155.041\,G$ for the condensates used in \cite{Donley}. For the case of Figure~\ref{fig7} we have $\lambda=6.8/17.5$, $d=2.16\,\mu m$ and $k_{eff}(t)=0.5715$, and correspondingly $\ell_{B}=0.199\,\mu m$. Since the interaction range we found is $0.1277d=0.276\,\mu m$, after making the various lengths dimensionful again we notice that the dimensionless ratio $\ell d/\ell_{B}^{2}$ is approximately equal to 15.

For the case of Figure~\ref{fig8} we have $\lambda=2.175$, $d=1.37\,\mu m$ and $k_{eff}(t)=0.6865$, and correspondingly $\ell_{B}=0.199\,\mu m$. Since the interaction range we found is $0.3165d=0.434\,\mu m$, after making the various lengths dimensionful again we notice that the dimensionless ratio $\ell d/\ell_{B}^{2}$ is approximately equal to 15.

It seems then that the dimensionful length $\ell$ is proportional to $\ell_{B}^{2}/d$.

\vskip 0.3cm
\vskip 0.3cm

{IV. \bf Conclusions}

The remnant condensate observed after the collapse of attractive Bose Einstein condensates has been a puzzle for some time, because the conventional Gross Pitaevskii formalism cannot readily account for its existence and its longevity. There have also been difficulties in its theoretical description because of its dissipative origin and of the abrupt and delayed onset of the collapse. The complex terms that had to be included in the extended Gross Pitaevskii equation seemed furthermore to make impossible the variational study of this equation.

In our paper we have addressed all these issues. By including nonlocal interaction terms in the action, we make the existence of the remnant inevitable and unavoidable. The collapse is understood now simply as the disappearance of the large width anisotropic condensate and its evolution to a narrow width spherical condensate.  This evolution necessarily reduces the number of atoms. This reduction necessitates however the inclusion of complex dissipative terms in the action. We presented a real action that results in a field equation incorporating the desirable complex terms. These terms contain explicitly the delayed onset of the collapse, and they are time-dependent, so that after the elapse of enough time they vanish, leading to a constant again number of atoms for the remnant. The reality of the action enables us to perform various variational calculations. These demonstrate that even though the remnant has eventually a constant number of atoms, it performs persistent and huge oscillations at frequencies $2\omega_{\rho}$ and $2\omega_{z}$. 

The proposed action can be used to perform variational calculations on any aspect of the behavior of the remnant condensate.

{\bf Acknowledgement}

We are grateful to Dr. Neil Ryan Claussen for his very helpful clarifications regarding the experimental results.


\begin{thebibliography}{99}
%
\bibitem{Donley}
E.A. Donley, N.R. Claussen, S.L. Cornish, J.L. Roberts, E.A. Cornell, C.E. Wieman, Nature {\bf 412} (2001) 295.
%
\bibitem{Ueda}
Masahito Ueda and Hiroki Saito, J. Phys. Soc. Jpn. {\bf 72} (2003) Suppl. C pp. 127-133
%
\bibitem{Savage}
C.M. Savage, N.P.Robins and J.J. Hope, \pra {\bf 67} (2003) 014304.
%
\bibitem{GPusers}
Antonios Eleftheriou and Kerson Huang, \pra {\bf 61} (2000) 043601; Yu. Kagan, A.E. Muryshev and G.V. Shlyapnikov, \prl {\bf 81} (1998) 933; Victo S. Filho, T. Frederico, Arnaldo Gammal and Lauro Tomio, \pre {\bf 66} (2002) 036225; L. Bergé and J. Juul Rasmussen, Physics Letters A {\bf 304} (2002) 136.
%
\bibitem{Parola}
A. Parola, L. Salasnich and L. Reatto, \pra {\bf 57} (1998) R3180; V.M. Perez-Garcia,  V. V. Konotop,  J.J. Garcia-Ripoll, \pre {\bf 62} (2000) 4300; F. Maucher, S. Skupin and W. Krolikowski, Nonlinearity {\bf 24} (2011) 1987; Fernando Haas and Bengt Eliasson, J. Phys. B: At. Mol. Opt. Phys. {\bf 51} (2018) 175302.
%
\bibitem{Gammal}
A. Gammal, T. Frederico and Lauro Tomio, \pra {\bf 64} (2001) 055602.
%
\bibitem{Bateman}
H. Bateman, Physical Review {\bf 38} (1931) 815.
%
\bibitem{Claussen}
For the most recent experimental results of the JILA group: Neil Ryan Claussen, PhD thesis, University of Colorado 2003.
%
\bibitem{Altin}
P. A. Altin, G. R. Dennis, G. D. McDonald, D. Doring, J. E. Debs, J. D. Close, C. M. Savage, and N. P. Robins, \pra {\bf 84} (2011) 033632.
%
\bibitem{Calzetta}
E. A. Calzetta and B. L. Hu, \pra {\bf 68} (2003) 043625.
%
\bibitem{Andreas}
S. Theodorakis, A. Hadjigeorgiou, J. Phys. B {\bf 50} (2017) 235301.
%
\bibitem{error}
N. R. Claussen, S. J. J. M. F. Kokkelmans, S. T. Thompson, E. A. Donley, E. Hodby, and C. E. Wieman, \pra {\bf 67} (2003) 060701R.




                        \end{thebibliography}
\end{document}